\documentclass[preprint]{aastex}
\usepackage{epstopdf}

\shorttitle{Tidal and Magnetic Interactions between a Hot Jupiter and
its Host Star}
\shortauthors{Chang et al.}

\begin{document}


\title{Tidal and Magnetic Interactions between a Hot Jupiter and
its Host Star in the Magnetospheric Cavity of a Protoplanetary
Disk}


\author{S.-H.  Chang\altaffilmark{1,2} and P.-G. Gu\altaffilmark{1} }


\and

\author{P. H. Bodenheimer\altaffilmark{3}}

\altaffiltext{1}{Institute of Astronomy and Astrophysics, Academia Sinica, Taipei 10617,
Taiwan, R.O.C.}
\altaffiltext{2}{Graduate Institute of Astrophysics, National Taiwan University, Taipei 10617, Taiwan, R.O.C.}
\altaffiltext{3}{UCO/Lick Observatory, University of California, Santa Cruz, CA 95064, U.S.A.}


\begin{abstract}
We present a simplified model to study the orbital evolution of a young hot Jupiter
inside the magnetospheric cavity of a proto-planetary disk. The model takes into
account the disk locking of stellar spin as well as the tidal and magnetic
interactions between the star and the planet. We focus on the orbital evolution
starting from the orbit in the 2:1 resonance with the inner edge of the disk,
followed by the inward and then outward orbital migration driven by the tidal and
magnetic torques as well as the Roche-lobe overflow of the tidally inflated planet.
The goal in this paper is to study how the orbital evolution inside the
magnetospheric cavity depends on the cavity size, planet mass, and orbital
eccentricity. In the present work, we only target the mass range from 0.7 to 2
Jupiter masses. In the case of the large cavity corresponding to the rotational
period $\approx 7$ days, the planet of mass $>1$ Jupiter mass with moderate initial
eccentricities ($\gtrsim 0.3$) can move to the region $< 0.03$ AU from its central star in
$10^7$ years, while the planet of mass $<1$ Jupiter mass cannot. We estimate the
critical eccentricity beyond which the planet of a given mass will overflow its
Roche radius and finally lose all of its gas onto the star due to runaway mass
loss. In the case of the small cavity corresponding to the rotational period
$\approx 3$ days, all of the simulated planets lose all of their gas even in
circular orbits. Our results for the orbital evolution of young hot Jupiters may
have the potential to explain the absence of low-mass giant planets inside $\sim
0.03$ AU from their dwarf stars revealed by transit surveys.

\end{abstract}





\section{Introduction}

The number of discovered extrasolar planets has been increasing
quickly\footnote{
See the Extrasolar Planets Encyclopaedia (http://exoplanet.eu) for
detail.}
since the first extrasolar planet, 51 Pegasi b, was detected
around a solar-type star \citep{Mayor95}. 51 Pegasi b is the
prototype of a class of exoplanets known as ``hot Jupiters"
because they are generally Jupiter-mass planets located $\lesssim
0.1$ AU from their parent stars and therefore their equilibrium
temperatures are much higher than that of Jupiter. The high
temperature of the protoplanetary disk at the orbital radii where
they are now provides an environment extremely different from
where giant planets are expected to form via either the
core-accretion model \citep{Pollack96,KI02} or the gravitational
instability model \citep{Gammie01,Boss04,Rafikov05}. One of the
most commonly adopted solutions is that hot Jupiters have formed
at larger orbital radii in the protoplanetary disk and then moved
inward to their current locations via the disk-planet interactions
\citep[e.g.][]{Lin96}.

The magnetic fields of Classical T Tauri stars (hereafter CTTS) are typically
strong enough to truncate the inner regions of protoplanetary disks to the
corotation radius, where the Keplerian angular velocity of the disk is the same as
the stellar-spin angular velocity, and create an inner magnetospheric cavity.
Besides, the migration due to the planet-disk interaction is expected to slow
dramatically once a young hot Jupiter passes the inner disk edge \citep{Lin96}. The
pile-up population of hot Jupiters at $\sim 0.04$ AU from their solar-type parent
stars revealed from radial-velocity searches \citep{Gaudi05} may be attributed to
the existence of the cavity \citep[][and references therein]{Carr07}.

On one hand, an attempt has been made by \citet{Setiawan08} to search for a hot
Jupiter around a CTTS,
although \citet{Huelamo08}
substituted an explanation of a cool stellar spot for an orbiting hot Jupiter.
Future observations will provide more evidence for planets of early ages. On the
other hand, the simulated orbital evolution of giant planets as they migrate into
the cavity was carried out by \citet{Rice08}. Their results suggest that high-mass
hot Jupiters would move closer to their central stars and even be destroyed because
planet's entry into the magnetospheric cavity results in a growth in orbital
eccentricity $e$. Although the eccentricity excitation is subject to the uncertain
properties such as disk mass and viscosity, this model may provide a mechanism to
pump up the eccentricity of a hot Jupiter inside the cavity and in turn affect the
orbital evolution of the planet via the tidal interactions between the star and
planet. In addition, the magnetic interactions between the star and planet in the
magnetospheric cavity would be important as well, which may be an upscale analogy
to the magnetic interactions between Jupiter and the Galilean satellites
\citep[][and references therein]{Zarka07}.

The coupled tidal evolution of the radius and orbit of hot Jupiters in the absence
of a disk has been investigated \citep{Gu03,Mardling07,IB09}. To obtain noticeable
influence on $R_p$, the tidal heating is assumed to be deposited deep in the planet
in these models \citep{Gu04}. The purpose of \citet{Mardling07} and \citet{IB09} is
to explain the large radii of some transiting hot Jupiters using tidal heating. On
the other hand, \citet{Gu03} showed that for modest eccentricities, young hot
Jupiters of radius $\approx 2$ Jupiter radii at $< 0.04$ AU swell beyond two
Jupiter radii and their internal degeneracy is partially lifted\footnote{For
comparison, the initial orbits and eccentricities of a young hot Jupiter in
\citet{IB09} are larger. Therefore the tidal heating in their model can proceed at
a modest rate to attain the large $R_p$ on the timescale of $\sim$ Gyrs.}.
Thereafter, their thermal equilibria become unstable and they undergo runaway
inflation until their radii exceed the Roche radius. Then Roche lobe overflow
ensues. Under the assumption that the overflowing gas immediately loses its orbital
angular momentum to the planet and plunges into the central star, these mass-losing
planets migrate outwards, such that their semi-major axes and Roche radii increase
while their mass, eccentricity, and tidal dissipation rate decrease until the mass
loss is quenched. Of course, including a disk would change the orbital evolutions
of Roche-lobe-overflowing planets, as has been calculated by \citet{Trilling98} who
do not consider the tidal expansion of $R_p$ and the magnetospheric cavity though.

As more and more hot Jupiters located $\lesssim 0.04$ AU have been discovered by
transit surveys for recent years, it is noteworthy that almost no giant planets of
mass $<1$ Jupiter mass ($M_j$) have been detected within $\sim 0.03$ AU from their solar-type parent
stars \citep[see Figure \ref{correl}; also see][and references therein]{US07}. This
puzzle may be attributed to the subsequent evolutions of hot Jupiters after they
enter the magnetospheric cavity of the protoplanetary disk. Since thermal
evaporation \citep[e.g.][]{DW09} has not been undoubtedly identified to be the
major cause \citep[][and references therein]{Murray09}, we shall discuss the
potential implication of our model to explain the observation in the end of the
paper.

For the above reasons, in this work we concentrate on the orbital evolution of a
young hot Jupiter inside the magnetospheric cavity of a protoplanetary disk by
taking into account the tidal and magnetic interactions between the star and
planet. To restrict to the problem without significantly involving the planet-disk
interaction for simplicity, we consider the evolution after the hot Jupiter enters
the 2:1 orbital resonance with the inner edge of a disk. The initial eccentricity,
which is expected to be excited in the cavity before the planet migrates to the
resonance location, is treated as one of the free parameters of the problem.
The equations for the orbital and
interior evolution of a hot Jupiter inside a magnetospheric cavity
are described in \S2. The results are presented in \S3, and are
summarized and discussed in \S4.

\section{Model Description}

\begin{figure}
$$
\scalebox{0.4}{\includegraphics{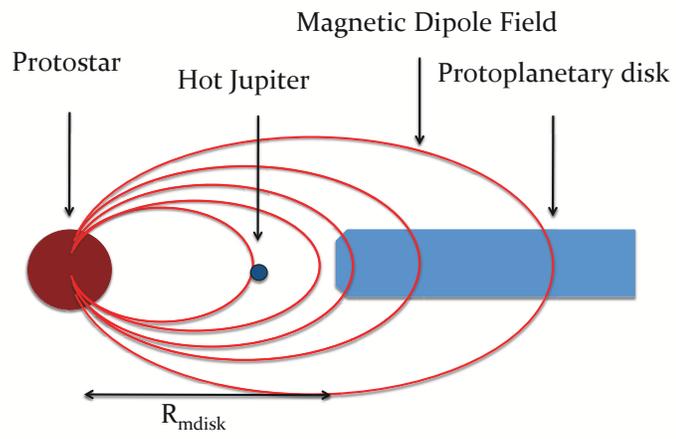}}
$$
\caption{A schematic illustration of our simple model.}
\label{model}
\end{figure}

Figure \ref{model} illustrates our toy model that shows a hot Jupiter orbiting a
CTTS in a magnetospheric cavity of a protoplanetary disk. The magnetic dipole
fields of a protostar truncate the inner part of the protoplanetary disk at the
disk radius $R_{mdisk}$, which is the radius of the cavity. The initial orbit of
the planet is set to lie at the 2:1 orbital resonance with the inner edge of the
disk, which allows us to ignore the tidal interactions between the disk and the
planet \citep{Lin96,Rice08}. The semi-major axis $a$ of this resonant orbit is
denoted as $a_{2:1}$. However, when the inner edge of the disk extends inwards for
any reasons in our model such that $a>a_{2:1}$, the planet is artificially moved
inwards as well to maintain the 2:1 orbital resonance with the inner edge of the
disk.
Moreover, we assume that there is no
magnetic linkage between the disk and the planet as illustrated in
Figure \ref{model} and therefore neglect magnetic interactions
between them. In other words, the planet's orbit evolves due to
tidal and magnetic interactions with the central star, whereas the
star interacts with both the planet and the disk. Therefore, this
section consists of two parts, namely the magnetic and tidal
effects, to describe the model in detail. \S2.1 and \S2.2 describe
the star-disk and star-planet magnetic interactions, which leads
to disk locking and contributes to the orbital evolution of the
planet, respectively. \S2.3 and \S2.4 then describe the tidal
interactions between the CTTS and the planet. The circularization
of the planet's orbit is due to tides on both the planet and the
star. The thermal inflation of the planet is due to tides on the
planet excited by the star. On the other hand, the tides on the
star raised by the planet affect the evolution of the planet's
orbit and the stellar spin. Finally, we summarize the key formulae
for our model in \S2.5.

\subsection{Star-Disk Magnetic Interaction}

There have been a number of models for the star-disk interactions
to account for the rotation of CTTSs as well as star and jet
formation \citep[e.g.][]{Shu94,MP05,Lovelace08}. In addition to
the closed field lines connecting from the protostar to the disk
to allow for funnel accretion, the open field lines may diverge
primarily from the inner edge of the disk \citep[i.e. X-wind
model, see][]{Shu94} or emanate largely from the star and the disk
\citep[i.e. stellar and disk wind model, see][]{MP05}.
Furthermore, the stellar field topology may be in the complex form 
of multipole \citep{Lovelace08} or tilted dipole relative to the stellar 
spin \citep{Laine}. In this work, we ignore these complications and 
adopt the simple non-tilted dipole configuration of stellar fields that 
thread through the protoplanetary disk as has been illustrated in Figure 
\ref{model}.

To calculate the magnetic torque $T_{mag}$ on the star due to the
magnetic linkage to the disk, we adopt the approach by
\citet{Armitage96} in which the magnetic torque acting on an
annulus of the disk is due to the azimuthal twisted field lines
$B_\phi$ caused by the difference between the stellar spin and the
orbital motion of the annulus \citep{LP92}. Hence, the torques
integrated from the disk inner edge at $R_{mdisk}$ to infinity
contribute to the net magnetic torque between the star and the
disk:
\begin{equation}T_{mag}=\frac{\mu^2}{3}(R_{mdisk}^{-3}-2R_c^{-3/2}R_{mdisk}^{-3/2}),\end{equation}
where $\mu=B_\ast R_\ast^3$ is the stellar dipole moment, $B_*$ is
the magnetic field strength at the stellar surface, $R_*$ is the
stellar radius, $R_c=(GM_\ast/\Omega_\ast^2)^{1/3}$ is the
corotation radius, $\Omega_*$ is the stellar spin rate, and $M_*$
is the stellar mass. We employ the same scaling law as used in
\citet{Armitage96} to specify the strength of stellar magnetic
field:
\begin{equation}
B_*=B_0 \left( {4{\rm days} \over 2\pi/\Omega_*} \right),
\end{equation}
where the scaling factor $B_0$ is a constant indicative of the
surface magnetic strength for the stellar spin period of 4 days.

The magnetospheric radius $R_{mdisk}$ is determined by the balance
between the stellar magnetic stress and the disk Reynolds stress
(Starczewski et al. 2007, and references therein):
\begin{equation}R_{mdisk}=\eta\left(\frac{B_\ast^4R_\ast^{12}}{GM_\ast\dot{M}_{disk}^2}\right)^{1/7},
\label{eq:R_mdisk}
\end{equation}
where $\dot{M}_{disk}$ is the disk accretion rate and $\eta$ is a dimensionless
factor of order unity. For a star with an aligned dipole field, $0.5\leq\eta\leq
1.0$. In addition, the mass transfer rate throughout the disk is assumed to be
comparable to an observational inferred accretion rate from protostellar disks onto
CTTSs, which decays with time $t$ over the timescale $\sim 10^7$ years and can be
fitted by
\begin{equation}\dot{M}_{disk}=10^{-4}\frac{M_\oplus}{\mathrm{yr}}\left(\frac{t}{10^7\mathrm{yr}}\right)^{-3/2},
\label{eq:Mdot}
\end{equation}
where  $M_\oplus$ is the Earth mass (Ida \& Lin 2004).

The equation for the spin evolution of the protostar is then given
by
\begin{equation}\frac{d\Omega_\ast}{dt}=-\frac{1}{I_\ast}(\dot{I_\ast}\Omega_\ast-T_{disk}).
\label{equation15}
\end{equation}
In the above equation, $\dot I_*$ is the moment inertia of the CCTS
and $T_{disk}=T_{mag}+T_{acc}$ is the net torque arising from the 
presence of the disk, where $T_{acc}$ is the torque spinning up the 
star due to the disk accretion:

\begin{equation}T_{acc}=\dot{M}_{disk}R_{mdisk}^2\Omega_{disk},
\label{equation17}
\end{equation}
with $\Omega_{disk}$ being the Keplerian angular speed at the radius
$R_{mdisk}$ \citep{Armitage96,MP05}.
$R_*$ and $\dot I_*$ are obtained from the interior structure of the 
contracting CTTS modeled by \citet{Armitage96}, who modified the 
Eggleton code \citep{Pols} for the pre-main sequence evolution.

When the star rotates faster than the inner edge of the disk (i.e.
$\Omega_* > \Omega_{disk}$), the mass accreting onto the star
should be ceased by the centrifugal force and $T_{acc}$ is
therefore set to be zero in our model. Meanwhile, the star
experiences immediate magnetic braking as a result of its strong
fields, leading to disk locking of the stellar spin.  Disk locking
makes the star and the inner disk edge corotate again and resumes
the accretion. Therefore precisely speaking, the disk density near
the inner edge of the disk, $R_{mdisk}$, and $\dot M_{disk}$
should fluctuate with time. In practice, we have simply employed
equations (\ref{eq:R_mdisk}) and (\ref{eq:Mdot}) to implement the disk
properties rather than modelling the detailed disk structure near
$R_{mdisk}$. This is certainly a limitation of our model. With
regard to the overall disk accretion rate, as we shall see in \S3,
the fluctuation of the balance between the mass accretion and the
magnetic braking (i.e. disk locking) is rather fast compared to
the timescales of interest, meaning that equation~(\ref{eq:Mdot}) is a
proper approximation to describe the average mass accretion rate
in our problem.

\subsection{Star-Planet Magnetic Interaction}

The torque on the planet due to magnetic linkage to the star is
modelled as the magnetic stress ($B_z B_\phi/4\pi$) multiplied by
the cross-sectional area of the planet's magnetosphere and the
distance $r$ between the planet and its parent star. It is assumed
that the planet has a magnetosphere in which the gas behaves like
plasma. In addition, the magnetosphere is assumed to be as large
as the planet's size, so the area of the planet interacting with the
stellar field is $2\pi R_p^2$. Applying the star-disk magnetic
interaction described in \S2.1 to the star-planet case, we obtain
the magnetic torque on the planet
\begin{equation}T_{planet}=\frac{B_\ast^2R_\ast^6R_p^2}{2}\left(\frac{1}{r^5}-\frac{\Omega_\ast}{r^5\dot{\phi}}\right),\end{equation}
where $\dot{\phi}$ is the orbital angular velocity.
The magnetic torque changes with the distance between the planet and the central star
while the planet moves in an eccentric orbit, so we employ the mean torque over one orbit period $P_{orbit}$.
That is,
\begin{eqnarray}
\int^{P_{orbit}}_0 T_{planet}\,dt
=\frac{B_\ast^2R_\ast^6R_p^2}{2}\int^{P_{orbit}}_0 \left(
\frac{1}{r^5}-\frac{\Omega_\ast}{r^5\dot{\phi}} \right) \,dt.
\label{equation3}
\end{eqnarray}
To carry out the above integral, we make the approximation that
the orbit does not vary significantly during one orbital period;
i.e., the orbital angular velocity $\dot{\phi}\approx
a^2n\sqrt{1-e^2}/r^2$ and $r\approx a(1-e^2)/(1+e\cos\phi)$, where
$a$ is the semi-major axis, $e$ is the eccentricity, and $n$ is
the mean motion of the mutual orbit of the planet and the star.
Then after integrating and being divided by $P_{orbit}$,  equation
(\ref{equation3}) gives an expression for the mean magnetic torque
on the planet:
\begin{equation} <T_{planet}>  =\epsilon\frac{B_\ast^2R_\ast^6R_p^2}{4}
\left[\frac{2+3e^2}{a^5(1-e^2)^{7/2}}-\frac{2\Omega_\ast}{a^5n(1-e^2)^2}\right].
\label{eq:T_planet}
\end{equation}
Here, we have introduced a coefficient $\epsilon$, where $0\le\epsilon\le1$, to
regulate the magnetic torque on the planet, a formulism similar to that applied to
the magnetic torques between the Galilean satellites and Jupiter (Zarka 2007). In
this regard, the energy dissipation associated with the magnetic torque can be
generally considered as the Poynting flux at the magnetopause scaled by $\epsilon$.
Hence, $\epsilon=1$ corresponds to the upper limit of the magnetic torque.

\subsection{Tides on the Planet}


The planet is assumed to have zero obliquity and zero orbital inclination. The
gravitational tides on the planet affect the evolution of the eccentricity and the
spin of the planet. For the sake of simplicity, we do not consider the influence of
the planet-star magnetic interactions on the spin and eccentricity of the planet.
Therefore, the rate of the eccentricity change of the orbit is due solely to
gravitational tides on the planet and the star. The rate contributed from the tidal
dissipation in the planet alone is given by $g_p$, which can be obtained from
(Mardling \& Lin 2002)
 \begin{equation}g_{p,\ast}=-\frac{81e}{{2Q'_{p,*}}}\left(\frac{M_{\ast,p}}{M_{p,\ast}}\right)\left(\frac{R_{p,\ast}}{a}\right)^5\left[f_1(e)n-\frac{11f_2(e)\Omega_{p,\ast}}{18}\right],
 \label{eq:TidalFrict}
 \end{equation}
where $g_\ast$ for the tidal dissipation in the star is also shown
and will be considered in \S2.4. In the above equation,
$f_1(e)=(1+15e^2/4+15e^4/8+5e^6/64)/(1-e^2)^{13/2}$ and
$f_2(e)=(1+3e^2/2+e^4/8)/(1-e^2)^5$. Moreover, $Q'_*$ and $Q'_p$
are the tidal quality factors of the star and the planet,
respectively.
All the uncertainties associated with the physical processes of
tidal heating are contained in the  $Q'$  values.
For a fiducial value, it has been suggested that the ${Q_p}'$  values inferred for
Jupiter from Io's orbital evolution is $5\times10^4<{Q_p}'<2\times10^6$ (Yoder
\&Peale 1981). With this ${Q_p}'$  value, orbits of planets with $M_p$ and $R_p$
comparable to those of Jupiter and with a period less than a week, are circularized
within the main-sequence life span of solar-type stars. $Q'_*$ has been inferred to
be $\sim 1.5\times 10^5$ for the very young binary stars \citep{Lin96}.

Since the synchronous timescale of the planet is much shorter than the
circularization timescale \citep{Gu03}, we simply assume that
the planet's spin equals the orbital rotation throughout the evolution. This
assumption may be invalid during the phase of the Roche-lobe overflow if the orbit
changes dramatically due to quick mass loss (see eq. (\ref{equation18}) below),
causing a large discrepancy between the planet spin and orbital motion. This means
that the tidal heating due to synchronization is unrealistically ignored during
this special phase in our model.

Moreover, the tidal process of orbital circularization shrinks the planet's orbit;
i.e.,
\begin{equation}\dot{a}=\frac{2e\dot{e}}{1-e^2}a.
\label{equation19}
\end{equation}
During the orbital circularization, there is substantial tidal heating within the
planet. The tidal heating of the planet may have contributed significantly to the
thermal budget that governs the planet's physical properties, including its radius
\citep{Bodenheimer01,Gu03,Gu04,Mardling07,IB09}.

Ignoring the rotation energy of the planet,
we can write the first law of the thermodynamics for a tidally heated planet as follows:
\begin{equation}\dot{E}_{tide}-\mathcal{L}=\dot{U}+\dot{W},
\label{equation6}
\end{equation}
where $\dot{E}_{tide}$ is the tidal heating rate (see Mardling \&
Lin 2002 for the expression), $\mathcal{L}$ is the intrinsic
luminosity from the photosphere of the planet, $U$ is the internal
energy, and $W$ is the gravitational potential energy. Suppose
that $U=q_UGM_p^2/R_p$ and $W=-q_WGM_p^2/R_p$ with the
coefficients $q_U$ and $q_W$ depending on the interior structure
of the planet. Then in the absence of mass loss (i.e.
$\dot{M}_p=0$), the above energy equation gives rise to the
expansion/contraction rate of the planet
\citep[cf.][]{Mardling02}\footnote{The internal energy is ignored
in \citet{Mardling02}}:
\begin{equation}\left(\frac{\partial R_p}{\partial t}\right)_{\dot{M}_p=0}=\frac{\dot{E}_{tide}-\mathcal{L}}{(\frac{d\ln q_U}{d\ln R_p}-1)\frac{U}{R_p}-(\frac{d\ln q_W}{d\ln R_p}-1)\frac{W}{R_p}} .
\label{equation7}
\end{equation}
Even though the expanding/contracting planet is not in the thermal
equilibrium, $q_U$, $q_W$, and $\mathcal{L}$ can be simply
expressed in terms of functions of $R_p$ and $M_p$ as long as the
timescales under consideration are $>$ the convective 
turnover time for an almost convective planet with the same solid-core
mass and a thin radiative envelope of the same opacity (i.e. an
almost polytropic interior, see Gu et al. 2003, 2004). In our
approach, the core mass of the planet is assumed to be
$19.4M_\oplus$, which is motivated by
the core accretion model for the formation of giant
planets (Pollack et al. 1996). However,
the thermal properties of the core are not modelled in our work.
$q_U(R_p)$, $q_W(R_p)$, and $\mathcal{L}(R_p)$ are
obtained only from certain different $M_p$
by fitting to the numerical data for the interior structures
of tidally inflated Jupiters described in \citet{Gu03}.
Subsequently, their values and derivatives with respect to $R_p$
for any desired $M_p$ can be interpolated from these available
fitted functions.

Once the planet's radius exceeds its Roche radius $R_l$ (see the Appendix), the
mass loss ensues. During the planet overflow phase, the $R_p$ adjustment after
losing mass plays an important role in determining the subsequent evolution.
$R_p$ changes with $M_p$ as well according to the equation of state of the planet
even though the entropy of the planet does not change. Since the interpolated data
derived from our interior structure code is unable to deal with the processes that
are as quick as the adiabatic mass loss, we adopt the following simple methodology.
Assuming a polytropic structure of a planet, the adiabatic mass-radius exponent
$\alpha\equiv(\partial \ln R_p/\partial \ln M_p)_s$ can be estimated from the
polytropic index of the planet \citep{Gu04}. $\alpha$ can then be related to the
gravitational potential energy. It can be easily shown that
$\alpha=(3-4q_W)/(3-2q_W)$. For simplicity we shall use this relation to estimate
$\alpha$ from $q_W$ for a Roche-lobe filled planet, albeit these polytropic
relations are derived for an isolated gravitationally bound body. Therefore, the
complete expression of $\dot{R_p}$ in the presence of the mass loss through the
Lagrangian 1 (L1) point should read
\begin{equation}\dot{R_p}=\left(\frac{\partial R_p}{\partial t}\right)_{\dot{M}_p=0}
+\alpha \left(\frac{R_p}{M_p}\right) \dot{M}_p, \label{eq:Rdot}
\end{equation}
where we shall adopt the formalism in \citet{Kolb90} to estimate $\dot M_p$ (see
\S2.5 for the detail). In the case of a giant planet with large radius such as a
young hot Jupiter, $\alpha$ is always negative. In addition, $|\alpha|$ increases
with $R_p$, because the planet degeneracy is being lifted when its density goes
down \citep{Gu04}.

Similar to the physical condition for the formation of the magnetospheric cavity of
the disk, planet's overflow may either form a ring/disk or funnel along the stellar
field lines onto the CTTS depending on whether $\dot M_p$ is strong enough to
dominate over the stellar fields. Therefore, we modify equation (\ref{eq:R_mdisk}) by
replacing $\dot M_{disk}$ with $\dot M_p$ to define the ``overflow cavity radius
$R_{oc}$" where the Reynolds stress of the overflowing gas is comparable to the
magnetic stress of the stellar fields. Thus, if this radius is larger than planet's
semi-major axis, where the overflow occurs, the overflowing gas will flow through
the magnetic field to the central star. The stellar spin will speed up under the
action of the torque from planet's overflow, $T_{\dot{M_p}}=\dot{M_p}(GM_\ast
a)^{1/2}$, which is an analogy to the spin-up torque due to the disk accretion
described in equation (\ref{equation17}). This leads to the evolution of the stellar
spin caused by the planet's overflow:
\begin{equation}\frac{d\Omega_\ast}{dt}=\frac{1}{I_\ast}\dot{M_p}(GM_\ast a)^{1/2}.
\label{equation18}
\end{equation}
On the other hand, if the overflow cavity radius is smaller than
the semi-major axis, the overflowing gas will form a ring/disk around
the star instead of directly funnelling along the field lines onto
the star.
In this stage, we assume that the total angular momentum of the
overflowing gas is immediately transferred to the orbit of the
planet, leading to the outward migration of the mass-losing
planet:
\begin{equation}\left(\frac{da}{dt}\right)_{overflow}=-\frac{2\dot{M_p}a}{M_p},
\label{eq:a_outward}
\end{equation}
and therefore no spin-up torque is exerted on the star.

The above equation for outward migration suggests that the mass-losing planet can
be detached from its Roche lobe as long as $R_p$ does not expand as fast as $R_l$.
Therefore, it is worthwhile to examine the linear stability of a Roche-lobe filled
planet against mass loss. The definition of $R_l$ from the Appendix and
equation (\ref{eq:a_outward}) give $\dot R_l/R_l=(-5/3)(\dot M_p /M_p)$.
Equation (\ref{eq:Rdot}) then indicates that the expansion rate of $R_p$ is $\approx
-\alpha \dot M_p/M_p$ provided that the thermal adjustment of $R_p$ is less
important than the adjustment of $R_p$ due to the adiabatic mass loss. Once the
planet's photosphere reaches the Roche radius (i.e. $R_p=R_l$), the above analysis
infers that the mass loss of a Roche-lobe filled planet is linearly unstable (i.e.
undergoes exponential growth) if $|\alpha|> 5/3$.


\subsection{Tides on the Star}

In addition to the tides on the planet induced by the star, the
planet also raises the tides on the star to circularize the orbit
and to lock the rotation of the star.

The circularization rate described in the preceding subsection has to be
complemented by the tidal dissipation in the star; namely,
\begin{equation}\frac{de}{dt}=g_p+g_\ast,
\label{eq:circularization}
\end{equation}
where $g_\ast$ has been displayed in equation (\ref{eq:TidalFrict}).

The process of the tidal locking of the CTTS leads to the exchange
of the angular momentum between the planet's orbit and stellar
spin. The stellar spin is assumed to be normal to the orbital
plane of the planet and the disk. Owing to this tidal effect, the
angular momentum of the stellar spin changes at the rate
\citep{Mardling02}
 \begin{equation}\dot{J}_\ast=\left(\frac{9J_0M_p}{M_t}\right)
 \left(\frac{f_4(e)\Omega_\ast-f_3(e)n}{2{Q_\ast}'(1-e^2)^{1/2}}\right)\left(\frac{R_\ast}{a}\right)^5,
\label{eq:J*dot}
 \end{equation}
where $J_0=M_p[GM_ta(1-e^2)]^{1/2}$ is the total angular momentum of the planet's
orbit, $f_3(e)=(1+15e^2/2+45e^4/8+5e^6/16)(1-e^2)^{-6}$, and
$f_4(e)=(1+3e^2+3e^4/8)(1-e^2)^{-9/2}$. The evolution of stellar spin associated
with equation (\ref{eq:J*dot}) is then given by
\begin{equation}\dot{\omega}_\ast=\left(\frac{M_p^2}{I_\ast M_t}\right)
\left(\frac{9n}{2{Q_\ast}'}\right)\left(\frac{R_\ast^5}{a^3}\right)[f_3(e)n-f_4(e)\Omega_\ast].
\label{eq:omega_*}
\end{equation}

\subsection{Equations for simulation}

After describing the various interactions separately, we list and
recapitulate the final equations incorporating these interactions
for solving the orbital evolution of a young hot Jupiter in a
magnetospheric cavity.

The rate of orbital circularization (i.e. eq. (\ref{eq:circularization}))
is given by
\begin{displaymath}\frac{de}{dt}=g_p+g_\ast.
 \end{displaymath}

The evolution of planet's size (i.e. eq. (\ref{eq:Rdot})) has
the form
\begin{displaymath}\dot{R_p}=\left(\frac{\partial R_p}{\partial t}\right)_{\dot{M}_p=0}
+\alpha \left( \frac{R_p}{M_p}\right) \dot{M}_p.\end{displaymath}
The first term on the right side is due to thermal adjustment,
whereas the second term is due to Roche-lobe overflow driven by an
adiabatic process that occurs much faster than the first term.

When $R_p > R_l$, the mass loss occurs in our simulation. This implicitly implies
that we neglect the mass loss from the optically thin atmosphere before the
planet's photosphere reaches its Roche radius. We also ignore the possible mass
loss through the L2 point \citep{Gu03}. The mass-loss rate of the planet is
estimated from the 1-D interior structure of the planet along with the formula in
Kolb \& Ritter 1990 (and see the definitions of the notations therein):
\begin{equation}\dot{M}_p=2\pi F_1(q)\frac{a^3}{GM_\ast}\int_{\Phi_L}^{\Phi_{ph}}F_3(\Gamma_1)\left(\frac{RT}{\mu}\right)^{1/2}\rho d\Phi.
\label{eq:M_dot}
\end{equation}

The orbital decaying rate of the planet is given by
\begin{equation}\frac{da}{dt}=\frac{2ae\dot{e}}{1-e^2}-\frac{2<T_{planet}>}{M_pan\sqrt{1-e^2}}+\frac{2a\dot{J}_\ast}{J_0}-\frac{2\dot{M_p}a}{M_p}.
\label{equation20}
\end{equation}
The first term on the right side represents the orbital decay due
to circularization (i.e. eq. (\ref{equation19})). The second term
is due to the magnetic torque on the planet (i.e. eq.
(\ref{eq:T_planet})) and the third term arises from tides on the
star (i.e. eq. (\ref{eq:J*dot})). The last term is included only
when the overflow occurs (i.e. eq. (\ref{eq:a_outward})).
Moreover, when $R_{mdisk}$ shrinks such that the 2:1 resonance
location lies in the disk, the planet is expected to slowly
migrate inwards \citep{Rice08}. However, when this happens, we
assume just for the sake of simplicity that $a$ is decreased
instantaneously to maintain the 2:1 resonance with the inner edge
of the disk.

The stellar spin evolution is determined by a number of factors:
\begin{equation}\frac{d\Omega_\ast}{dt}=-\frac{1}{I_\ast}\{\dot{I_\ast}\Omega_\ast-T_{disk}-<T_{planet}>+\dot{M}_p[GM_\ast a(1-e^2)]^{1/2}\}+\dot{\omega}_\ast.
\label{equation21}
\end{equation}
The first term in the curly bracket arises from the evolution of
the stellar moment of inertia during the pre-main-sequence phase
and the second term represents the torque exerted by the disk on
the star due to disk accretion and the magnetic linkage (i.e. eq.
(\ref{equation15})). The third and the fourth terms are due to the
magnetic linkages with the planet (i.e. eq. (\ref{eq:T_planet}))
and the overflowing gas from the overfilled Roche lobe (i.e. eq.
(\ref{equation18})), respectively. The last term
$\dot{\omega}_\ast$ results from the tides on the star (i.e.
eq. (\ref{eq:omega_*})).




\section{Numerical Results}

To solve the ODEs in the preceding section (eq. (\ref{eq:circularization}),
(\ref{eq:Rdot}), (\ref{eq:M_dot}), (\ref{equation20}), (\ref{equation21})), we
employ the 4th order Runge-Kutta method to study the inward migration of a young
hot Jupiter in the magnetospheric cavity due to tidal and magnetic interactions
with its CTTS. The simulations start from $t=10^5$ yr to $10^7$ yr. The planet is
not introduced to the orbit with $a=a_{2:1}$ until $t=7\times 10^5$ years, a
timescale comparable to the Type II migration time of a giant planet in a
proto-planetary disk \citep{LP85}.

In our model, $B_0$, $e$, and $M_p$ are parameterized to
investigate the various evolutions of the planetary migration in
the magnetospheric cavity. Besides, ${Q_\ast}'$, ${Q_p}'$ and $\eta$
are fixed as $3\times10^5$, $10^6$, and $0.75$ throughout the simulation.

\begin{figure}
$$
\scalebox{0.6}{\includegraphics{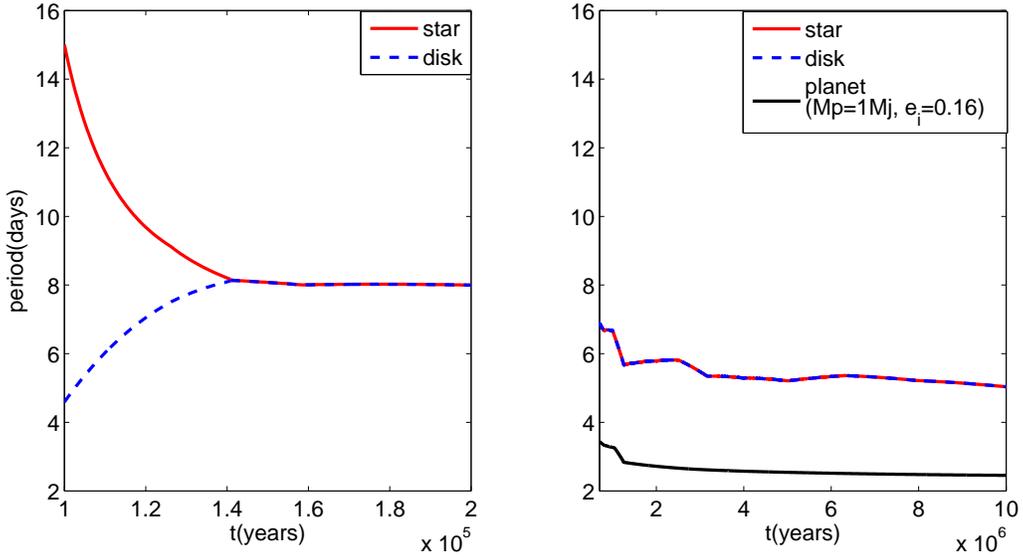}}
$$
\caption{The evolution of the stellar spin (red line), the rotational period of the
inner edge of the disk (blue dashed line), and the orbital period of the planet (black line). 
The left panel displays that the star's initial spin period (i.e., 15 days) at $t=10^5$
years is locked to 8 days by the disk in $\sim 4\times 10^4$ years. Afterwards, the
star corotates with the disk inner edge, and its spin rate is determined by the
disk accretion and the disk locking. The planet's orbital period is added in the
right panel to compare to the stellar spin period and the rotational period of the
inner edge after the planet is introduced at $t=7\times 10^5$ years.
The difference between the planet's orbital and stellar spin rates leads to the
planet's inward migration as shown by the black curve, while the star does not spin
up via the tidal effect and star-planet magnetic linkage due to the strong disk
locking.} \label{locking}
\end{figure}

The initial period of the central star is set to be 15 days at $t=10^5$ years (see
the left panel of Figure \ref{locking}). The figure shows the stellar spin will be
locked by the disk in a few $\times10^4$ years and therefore the initial spin is
insensitive to the tidal and magnetic interactions, which proceed on the timescales
of $\sim 10^{5-6}$ years. The final disk-lock period (i.e. the Keplerian period of
the disk inner edge) depends on the magnitude of the magnetic field of the central
star \citep{Armitage96}. We discuss two cases of cavity sizes corresponding to
$B_0=1500$G (\S 3.1) and $B_0=500$G (\S 3.2) separately. The disk with a big cavity
will lock the central star at the rotational period of $\sim 7$ days and that with
a small cavity will lock it at $\sim 3$ days. \citet{Herbst05} revealed the
peculiar bimodal distribution of periods for young stars in the Orion Nebula
Cluster. The distribution of periods at an age of about 1 Myr exhibit two peaks,
one located at a period of about 8 days and the other at a period of about 2-3 days
\citep{Bouvier07}. The disk-lock periods in our model approximately match the
observation.

We consider the planets of four different planetary masses $2M_j$, $1.5M_j$,
$1M_j$, and $0.7M_j$ to study their different evolutions of migration. From the
observations, the close-in transiting giant planets have radii ranging from
$\sim$0.8 $R_j$ to $\sim$1.7$ R_j$.
Thus 1.7$R_j$ could be considered as a lower limit of the radius of a young Jupiter
although some of the large planet's sizes have been postulated to be attributed to
tidal heating \citep{Miller09} or evaporation \citep{Baraffe}. Also, 1.7 $R_j$
corresponds to the theoretical radius of a young Jupiter (Marley et al. 2007). Thus
we fix the initial $R_p=1.7R_j$ for all the cases.

When $t=7\times 10^5$ yr and the planet is introduced to the initial location at
$a=a_{2:1}$, the star has been locked by the disk. The right panel of Figure
\ref{locking} shows the decrease in the periods of stellar spin and planet's orbit
before $t\lesssim 10^6$ years. This is caused primarily by the contraction of the
cavity as $R_*$ decreases (see eq. (\ref{eq:R_mdisk})). Orbiting interior to the
co-rotation radius, the planet then migrates inwards from the 2:1 resonance due to
faster revolution of the planet than the spin of the central star. This is
illustrated by the black curve during the time $t > 10^6$ years in the right panel
of Figure \ref{locking} where the planet with $1M_j$ and $e_i=0.16$ is used as an
example. Meanwhile, the star should spin up via the tidal and magnetic interactions
with the planet. However, the disk locking dominates the evolution of the stellar
spin as indicated by the overlap between the red solid and blue dashed curves in the
right panel of Figure \ref{locking}. In other words, the inner edge of the disk acts
as the sink of total angular momentum of the system in our model. In fact, the
rotational period of the disk inner edge fluctuates slightly from the spin period
of the star on short timescales. Nevertheless, these two periods are almost the
same over long timescales, justifying the argument given in \S2 that
equation (\ref{eq:Mdot}) is a proper approximation to describe the average mass
accretion rate and determine the cavity size in our problem.

As mentioned in \S 2.3, the planet will lose its mass if the planet's radius
exceeds the Roche radius. However, if the planet's mass decreases to
20 earth masses, the simulation will be forced to stop. This is
because the core mass of a giant planet is 19.4 earth masses in our model.

\subsection{The Large-Cavity Case}

A large cavity is opened up by the strong stellar magnetic fields with the scaling
factor $B_0=1500$ G.
In this case, the planet starts to migrate from $a_i\sim 0.045$ AU, which
corresponds to $a_{2:1}$ at $t=7\times 10^5$ years.

The migration evolutions of the planets with same masses will
differ as a result of different initial eccentricities. The
distance, over which the planet can migrate inward from the
initial position, is shortest when $e_i=0$ and increases as $e_i$
increases. This is because firstly, the circularization of a
eccentric orbit makes the orbit shrink. Secondly, the inflation of
a planet due to tidal heating enhances the star-planet magnetic
interaction. Also, the tidal torque on the star caused by the
planet is more significant as $e_i$ is larger. All of the above
factors accelerate the migration rate of the planet. When $e_i$
$\geq$
the critical eccentricity $e_c$,
the planet's radius will exceed its Roche lobe
and start to lose mass onto the central star.

We perform the calculations for three cases of $e_i$: (1) $e_i=0$;
(2) $e_i=e_c$; (3) $e_i=e_c-0.01$ for each planet mass. The
orbital evolutions are shown in Figure \ref{bigcavity} and the explanations
are presented below.

In the case of $e_i=0$ (solid lines in the figure), the results for different masses
are not significantly difference. Without an eccentric orbit to bring the planet
closer to the central star (i.e. the perihelion), the tidal and magnetic effects
are excited less effectively in a large distance between the planet and the central
star. Hence, planets with different masses initially in circular orbits migrate to
almost the same final positions, $a\sim 0.037$ AU at $t=10^7$ yr, as the cavities
contract and drive the planets to the 2:1 resonances.


When $e_i=e_c$ (dotted lines in the figure), the planets start to overflow
their L1 points and lose mass. As shown in the figure, their evolutions are
terminated before the end of the simulations ($t=10^7$ years), suggesting that all
of the planets are disrupted. The termination of the simulations occurs when the
discrepancy between $R_p$ and $R_l$ is so large during the overflow phase that
$\dot{M_p}$ is too large to be resolved in the simulation with the time step $=100$
years. This time step of 100 years is already shorter than the time step used for
the interpolated data derived from our interior structure code. When the runaway
solutions happen, $|\alpha|$ is always larger than 5/3, which is in agreement with
the linear instability of the mass loss. For the $0.7$ and $1M_j$ cases, the mass
loss rate is high. The overflow gas is able to form a ring/disk and hence the
planet migrates outward according to equation (\ref{eq:a_outward}) in our model. However,
once their overflows occur, $|\alpha|>5/3$ at the same time owing to their large
$R_p$. The radius adjustment is predominated by the adiabatic mass loss instead of
tidal inflation, leading to the runaway solutions to the ODEs.

On the other hand for the $1.5M_j$ and $2M_j$ cases, they migrate fast to the
locations closer to the central stars owing to the stronger tides raised by these
massive planets on their central stars. In addition, these more massive planets are
more difficult to inflate by tidal heating due to their higher gravitational
binding energies and cooling rates \citep{Gu03,Gu04}. Hence, when they overflow
their Roche lobes, $R_p$ is not too large and therefore $|\alpha|<5/3$, leading to
the stable overflow. Although the overflowing gas initially funnels to the central
star, the overflow rate becomes substantially high shortly such that $R_{oc}<a$.
Consequently the massive planets have undergone outward migration almost since the
overflow began. The outward migration is primarily driven by inward migration due
to the tidal and magnetic interactions
rather than tidal inflation. The overflow finally turns to be unstable when
$|\alpha|$ increases to 5/3 as $a$ and therefore $R_p$ increases. The evolution
then stops and the planet loses all its gas to the central star. During the phase
of the fast outward migration, the planet's spin may become unlocked with its
orbital motion. Despite not being included in our model, the tidal heating
associated with the asynchronism would speed up the planet inflation and lead to
the same catastrophic fate of the planet.
Comparing these cases for different initial $M_p$, massive planets require higher
$e_c$ thus more tidal heating to be inflated against their greater gravitational
binding energies and cooling rates. $M_p=2M_j$ requires $e_c=0.35$ and destructs at
$a\sim 0.02$ AU, while $M_p=0.7M_j$ needs $e_c=0.22$ and destructs at $a\sim 0.035$
AU.


Now we turn to the case for $e_i=e_c-0.01$ (i.e. $e_i$ is slightly
smaller than $e_c$) to study the conditions in which
the planet of a given mass can migrate inward the farthest and still survive in the
end of the simulation. As can be expected from the previous results for $e_i=e_c$,
the dashed curves in the figure show that a more massive planet undergoes faster
migration due to the stronger tidal torque. Thus, the planet with $M_p=2M_j$ and
$e_i=0.34$ can arrive at $a\sim 0.024$ AU, while the planet of $M_p=0.7$ and 1
$M_j$ can not migrate to the region $<0.03$ AU. Furthermore, a low mass planet is
more easily inflated by tidal heating, giving rise to the relatively stronger
magnetic interaction (see eq. (\ref{eq:T_planet})). This explains why the planet of
$0.7 M_j$ migrates slightly farther in than the planet of $1M_j$ in the end of the
simulation.


In equaiton (\ref{eq:T_planet}), the coefficient $\epsilon$ regulates the magnetic torque
on the planet, and the upper limit $\epsilon=1$ has been applied to obtain the
above results. Here, we employ $\epsilon=0$ to study the orbital evolutions in the
absence of the magnetic torque. In comparison to the previous orbital evolutions
shown in Figure \ref{bigcavity}, Figure \ref{eps0} shows the lower migration rates because
of the lack of the star-planet magnetic interaction for the $M_p=$0.7, 1, and
1.5$M_j$ cases. However, the stellar tides induced by the $2M_j$ planet are far
dominant over the magnetic interaction, so the massive planet still migrates to
$a\sim 0.016$ AU. That position is even farther inward than $a\sim 0.024$ AU for
the $\epsilon=1$ case as the planet with even higher $e_i=0.38$ can still survive
in the absence of the magnetic torque. Moreover, Figure \ref{eps0} confirms the
previous subtle result about the dependence of the magnetic torque on $R_p$: the
slightly more inward migration of the planet of $0.7M_j$ than $1M_j$ does not
reappear because these exists no magnetic torque to speed up inward migration for a
lighter planet of a larger inflated radius.
As to the cases for $e_i=e_c$, the simulated planets overflow their Roche radii and
perish in the same way as in the $\epsilon=1$ cases; namely, $|\alpha|$ of various
cases finally is $>5/3$ and the mass loss becomes runaway. For the cases of
$0.7M_j$, $1M_j$, and $1.5M_j$, $|\alpha|>5/3$ once their overflow occurs.
On the other hand, after reaching the Roche lobe, the planet of $2M_j$ migrates
outward for a period of time due to stable overflow and inward migration driven by
the tidal torque. It is finally destroyed due to the large $|\alpha|$ that
increases with $R_p$ as the mass-losing planet moves outwards.


\subsection{The Small-Cavity Case}
\begin{figure}
$$
\scalebox{0.5}{\includegraphics{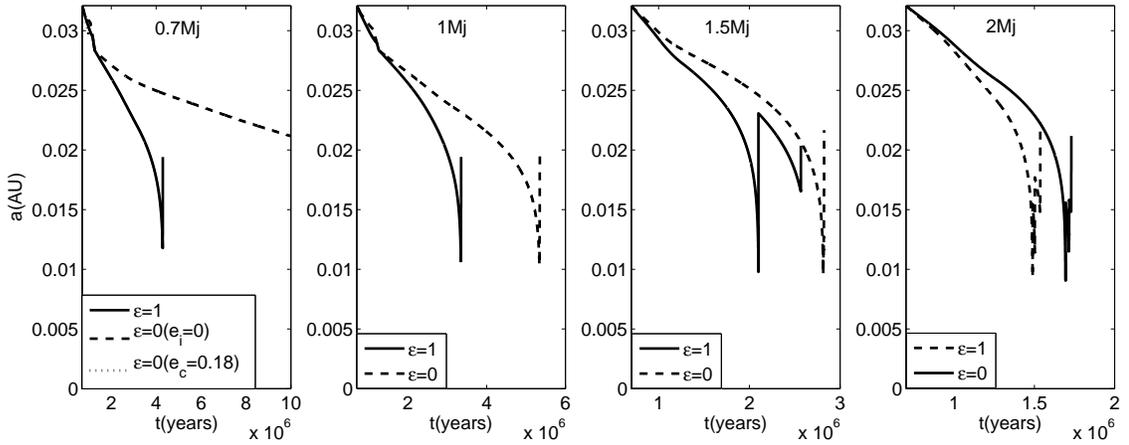}}
$$
\caption{Comparision between the cases with the magnetic torque in full strength
($\epsilon=1$ plotted in sold line) and without the magnetic torque ($\epsilon=0$
plotted in dashed line) for the orbital evolutions of the planets of differernt masses with
$e_i=0$ in the small cavities. Note that since the planet of $0.7M_j$ with $e_i=0$
survives up to the end of the simulation at $t=10^7$ yr, its $e_c$ is estimated and
the corresponding evolution is also shown (dotted line) in the leftmost panel. }
\label{eps0sc}
\end{figure}

The weaker stellar fields due to the smaller scaling factor $B_0=500$ G form a
smaller magnetospheric cavity, so the initial position of the planet is much closer
to the central star; namely, $a_i\sim0.032$ AU. The planet and the star exert
remarkable tidal forces on each other in such short distances, and the magnetic
torque on the planet is not negligible as well. $R_l$ is getting smaller while the
planet migrates inwards. As soon as $R_l<R_p$, the overflow occurs, leading to the
destruction. Consequently, even $e_i=0$ (i.e., no tidal heating in the planet), all
the planets destruct irrespective of their mass. If $M_p$ is bigger, the migration
is faster due to the stronger tidal torque. Therefore, the more massive the planet
is, the quicker the planet destructs, as shown in Figure \ref{eps0sc} for
$\epsilon=1$.

The orbital evolutions of the the $1M_j$ and $2M_j$ cases are characterized by the
alternating out/inward orbital migrations. Once the overflow occurs, the large
expansion of $R_l$ due to the high mass loss rate dominates over the adiabatic
inflation, hence leading to the fast outward migration. As a result, $R_p$
temporarily detaches from $R_l$ for a while until the planet moves in and the next
Roche overflow occurs again. Finally, the planet will lose all its mass when the
overflow occurs accompanied with high $|\alpha|$ (i.e. $>5/3$).

In the absence of the magnetic torque (i.e. $\epsilon=0$), the tides raised by
massive planets on the stars still drive significant inward migration of the
planets. Figure \ref{eps0sc} shows that shutting down the magnetic torque cannot
change the outcome of planet destruction for $M_p \geq 1M_j$. However, the planet
of $0.7 M_j$ can survive in $10^7$ years because the inward migration driven by the
weak tidal torque from the low mass planet is slow. Therefore, we estimate
$e_c=0.18$ for the planet to perish and show the evolution in the leftmost panel of
Figure \ref{eps0sc}.

\section{Summary and Discussion}

We construct a simple model to study the orbital evolution of a young hot Jupiter
in an eccentric orbit inside a magnetospheric cavity of a proto-planetary disk
around a CTTS.
For the sake of simplicity, we assume that the magnetospheric cavity of the
protoplanetary disk is truncated by a stellar dipole field. Through the magnetic
linkage between the central star and the disk, the rotational period of the star is
quickly locked by the inner edge of disk. We then introduce a young hot Jupiter at
$a_{2:1}$ and restrict ourselves to the orbital evolution of a hot Jupiter inside
the cavity.
Assuming that the planet behaves like a plasma, we apply the same formulism for the
star-disk magnetic interaction to that for the star-planet interactions. We adopt
the equilibrium-tide equations with $Q_*=3\times 10^5$ and $Q_p=10^6$ to model the
star-planet tidal interactions.
In so doing, our model focuses on the planet migration due to tidal and magnetic
interactions between the planet and the star, and does not consider any
interactions between the planet and the disk. However, as the size of the
magnetospheric cavity evolves, the planet is artificially pushed to $a_{2:1}$ if it
lies beyond $a_{2:1}$. We vary three parameters $B_0$ ($=500$ and 1500 G), initial
$M_p$ ($=0.7$, 1, 1.5, 2 $M_j$), and initial $e$ ($e_i=0$, $e_c-0.01$, and $e_c$)
in our simulations to investigate the fate of the planet migration under the
influence of the cavity size, planet mass, and orbital eccentricity.

Two sizes of magnetospheric cavities are considered to approximately match the
bimodal distribution of spin periods of young stars in the Orion Nebula Cluster.
The initial $a_{2:1}$ corresponding to the large (small) cavity opened up by the
stronger (weaker) stellar fields is $\sim 0.045$ AU ($\sim 0.032$ AU). In the case
of the large cavity ($B_0=1500$ G), the planets require nonzero $e_i$ to enhance
their tidal and magnetic interactions with their central stars and make significant
inward migration. The migration rate increases with $e_i$ for a given planet mass.
When $e_i$ is as large as the critical value $e_c$, $R_p$ exceeds $R_l$ and
the Roche overflow occurs. For planets with different masses, massive planets
($M_p>1M_j$) have higher $e_c$ (i.e. more intense tidal heating) than less massive
ones ($M_p\le M_j$) because massive planets are more difficult to thermally inflate
due to their greater gravitational binding energies and cooling rates. As a result,
stronger stellar tides raised by the planet with higher $M_p$ and $e_i$ allow for
faster migration. Overall, massive planets can therefore migrate further in than
less massive ones, as shown in the simulations for $e_i=e_c-0.01$. When $e_i=e_c$,
low-mass planets ($M_p=0.7$ and 1 $M_j$) inflate by tidal heating faster than
inward migration. Hence they overflow their large Roche lobes at large $a$,
resulting in the low density and therefore the runaway mass loss. On the other
hand, the high-mass planets ($M_p=1.5$ and 2 $M_j$) migrate inwards fast without
significant tidal inflation. In contrast to the low-mass planets, once these planets fill their
small Roche lobe at small $a$, their density is high enough that they can undergo
stable L1 overflow. During this phase, tidal and magnetic interactions instead of
tidal inflation drive them to lose mass, expand, and migrate outward. This stable
mass-loss phase proceeds until these outward migrating planets are large enough to
become unstable against mass loss. Finally, they lose all gas to their central
stars.

In the case of the small cavity ($B_0=500$ G), the simulated planets in circular
orbits all quickly migrate in due to the fierce tidal and magnetic interactions
until they stably overflow their Roche radii. After that, they can move outwards
due to the mass loss and move inwards again by the tidal and magnetic torques. As
their mass goes down and their degeneracy is lifted, all of these planets are
finally destroyed, suffering from the runaway mass loss as in the large-cavity
case.

To study the significance of the magnetic interactions between the star and the
planet in our model, we also simulate the cases by shutting down the interaction
(i.e. $\epsilon=0$). We found that the migration of less massive planets is
more sensitive to the magnetic interaction, which is enhanced by their easily
inflated radii. Setting $\epsilon=0$ makes the migration rate of low-mass planets
slower, as shown in the case for $M_p= 0.7M_j$ with $e_i=0.21$ in the large-cavity
case. By contrast, the inward migration of massive planets is less sensitive to the
magnetic interaction but more to the tidal interaction. As has been summarized in
the preceding paragraph, all planets in the small cavity destruct under both the
tidal and magnetic torques at such close-in initial positions from the stars even
if $e_i=0$. When $\epsilon=0$, only the planet of $0.7M_j$ can survive unless the
low-mass planet starts with an eccentric orbit with $e_c=0.18$.

The model presented here for the orbital evolution of young hot Jupiters is
different from the works by \citet{Gu03} and \citet{Trilling98}. In this work, we
employ simple models including the star-planet interactions and disk locking, which
are not considered by \citet{Gu03}. Furthermore, we take into account the planet
expansion due to the adiabatic mass loss and model the mass loss rate using the
approach in \citet{Kolb90}, whereas \citet{Gu03} ignore the radius adjustment to
the mass loss and use a free parameter to control the mass loss rate. Ignoring the
mass-radius relation during the Roche-lobe overflow phase as done in \citet{Gu03},
a Roche-lobe filled giant planet with $e_i=e_c$ continuously loses its mass via
tidal inflation and migrates outwards until the tidal heating rate is weaker than
the cooling rate, leading to a survived planet of lower mass. On the other hand,
\citet{Trilling98} consider young hot Jupiters moving inwards via the Type II
migration without a magnetospheric cavity and tidal inflation. In their work
including the mass-radius relation, once the planets of mass $< 3.36 M_j$ get close
enough to their central stars and therefore overflow their Roche lobes, they
migrate outwards and finally lose all their mass onto the star. In this regard, our
results are compatible with theirs. One of the major differences is that the mass
loss is driven by the Type II migration in the model by \citet{Trilling98}, while
in our work the mass loss is driven by the loss of orbital angular momentum via the
tidal and magnetic interactions with the parent star whose spin is almost locked by
the disk. Moreover, the planets of mass $\leq 2 M_j$ with $e_i<e_c$ survive in the
cavity during the CTTS phase in our model.

For the last few years, transit surveys have revealed the peculiar $M_p-a$
correlation for hot Jupiters: less massive hot Jupiters ($< 1$ Jupiter mass) are
almost absent within $\sim0.03$ AU from their parent dwarf stars of the ages $\sim$
Gyrs.
Our results show that during the CTTS phase, planets of $M_p\ge1M_j$ with modest
initial eccentricities ($e_i\gtrsim 0.3$) and sufficient magnetic interactions in a
large cavity can migrate to $a < 0.03$ AU, while planets of $M_p<1M_j$ cannot. More
specifically, given $e_i<e_c$, the planet of $2M_j$ can safely arrive at $a\lesssim
0.024$ AU and the planet of $1.5M_j$ can get to $a\lesssim 0.025$ AU in the end of
our simulations. Whether or not these planets can further migrate toward their
spun-down parent stars during the main-sequence phase depends on $Q'_*$, which is
not yet well understood.

\begin{figure}
$$
\scalebox{0.5}{\includegraphics{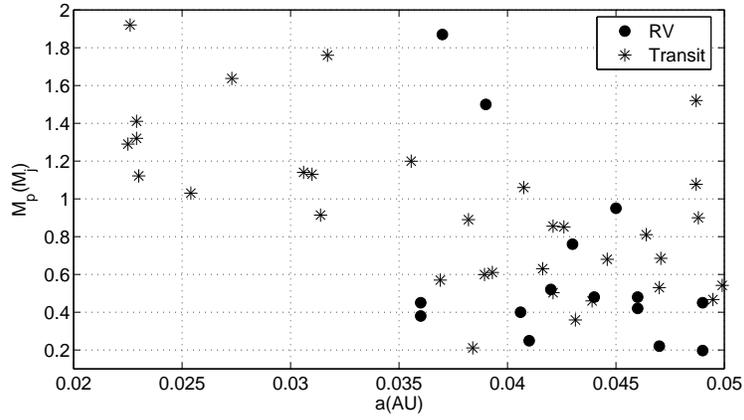}}
$$
 \caption{The semi-major axis vs. mass  relation for the hot Jupiters of mass between $30$
        Earth masses and $2$ Jupiter masses. The data are
adopted from the website http://exoplanet.eu/. The transiting exoplanets are shown
by the asterisks and the exoplanets discovered in radial velocity surveys are
denoted by the black circles. The exoplanets that are known to be in retrograde
orbits are excluded in the plot because their orbital orientations are unlikely to
be explained by the planet migration model as presented in this paper
\citep{Nagasawa08,Anderson09,Narita09}.} \label{correl}
\end{figure}

\citet{OL07} suggested that $Q'_*$ induced by hot Jupiters may become large (i.e.
$\gg 10^6$) as CTTSs evolve to main-sequence stars based on their dynamical-tide
model. As they and others \citep{Patzold02,Jiang03} pointed out, we would be
extremely fortunate to observe these very close-in hot Jupiters if the current
$Q'_*\sim 10^6$, implying that $Q'_*$ of the solar-type main-sequence host stars
may be quite large. However, it should be noted that these very close-in planets
were all discovered by transit surveys (see Figure \ref{correl}). Although transit surveys
overall prefer to detect close-in planets \citep{Gaudi05}, the detectability of a
hot Jupiter as a function of the semi-major axis of interest here (i.e. $a\lesssim
0.05$ AU) is extremely ill-defined as the various transit surveys are subject to
different observational limitations such as the transit depth, sampling rate, red
noise, and detection threshold \citep{Pont06}.
If $Q'_*$ indeed gradually becomes $\gg 10^6$ as CTTSs evolve to main-sequence
stars, therefore allowing the planets to move a little bit inward during the short
transition after $t=10^7$ years while halting the further planet migration for most
of the main-sequence phase, our model may have the potential to explain the absence
of low-mass hot Jupiters within $a\lesssim 0.3$ AU.

On the other hand, a large body of studies have suggested that
$Q'_\ast$ associated with the tidal dissipation induced by the hot Jupiters in
main-sequence dwarfs could be as small as $\sim 10^{5-6}$. As a result of the
efficient tidal dissipation, the orbits of the hot Jupiters can decay so
significantly that the planets may plunge into their host stars during the
main-sequence phase \citep{Jiang03,Jackson08}. This ``accretion" model may account
for the paucity of the extremely close-in hot Jupiters concluded from
radial-velocity surveys \citep[][also see Figure \ref{correl}]{Gaudi05}. In addition, the
relatively young ages of the extremely close-in hot Jupiters revealed by transit
surveys may also lend support to the model \citep{Jackson09}. Although the
planet-metallicity correlation independent of the spectral type does not seem to
favor the accretion model \citep{Fischer05}, excessive heavy elements in the
shallow convection zone of high-mass dwarfs may be able to sink down to the
radiative zone via the double diffusive effect, thereby eliminating the metallicity
enhancement and resolving the problem for the model \citep{Vauclair}. 
Furthermore, \citet{Pont09} found an excess spin of the host stars
of hot Jupiters, which may arise from tidal spin-up, although the uncertainty of
the stellar ages
and the limitation of a small number of samples demand more
detections in the future to confirm the conclusion.
The study by \citet{Jackson09} for the accretion model is carried out based only on
one-mass case (i.e. $1M_j$). If the accretion model is the main cause responsible
for the observed $M_p-a$ relation, it would be expected that the extremely close-in
giant planets of mass $<1M_j$ of younger ages should also be detected by transit
surveys.
The orbital decay timescale during the main-sequence phase due only to the 
tidal dissipation is given by $t_{tide} \approx (J_0)/(2\dot{J_*})$ (see equation(\ref{eq:J*dot})). 
In the case of a planet of 0.7$M_J$, the planet can survive in $\sim10^9$ yrs from 
the distance of $\sim0.03$AU at $t=10$ Myrs. In contrast, the orbit of a planet 
of $2M_J$ will decay on the timescale of $\sim10^7$ yrs from a distance of 
$\sim0.025$AU at $t=10$ Myrs. In other words, during the main-sequence phase, 
less massive planets excite weaker tides on their parent stars and therefore still stay outside 
$a\sim0.03$ AU, while the massive planets keep falling into the stars in the way as 
interpreted by \citet{Patzold02} and \citet{Jiang03}. The recently discovered hot Jupiter 
WASP-18b with an orbital period of 0.94 days and a mass of $10M_J$ may provide a 
contraint on $Q_*'$ of a main-sequence star in the next few years \citep{Hellier}.

In this paper, we have attempted to bring several physical components to the
context of the evolution of young hot Jupiters. Nonetheless, a large number of
assumptions have been made to simplify the physical processes included in our
model. The structure of the truncated disk is not modelled. How a planet migrates
to $a_{2:1}$ and how the eccentricity evolves during its entry to the cavity
\citep[e.g.][]{Rice08} are not addressed. The disk accretion rate is assumed to be
quasi-steady to determine the cavity sizes, while observations show otherwise
\citep[e.g.][]{Baraffe09}. The dipole configuration of the stellar fields
connecting to the disk ignores any types of winds even though the effect of disk
locking is introduced in a simplified manner to imitate the sink of the total
angular momentum in our model. The star-planet magnetic interactions are
parameterized to scale with the Poynting flux at the magnetopause of the planet
rather than appealing to any specific models, such as the dissipation due to the 
Alfv\'en waves and unipolar inductor \citep[e.g.][]{Zarka07} or 
induced by tilted stellar dipole fields (e.g. Laine et al. 2008; cf. Papaloizou 2007). The
values of $Q'_*$ and $Q'_p$ chosen for the calculations are quite arbitrary; we
have not explored a wide range of the values. The interior structure of an inflated
planet and the mass-radius relation of a hot Jupiter are inferred from the 1-D
numerical simulation \citep{Bodenheimer01,Gu03} even for a Roche-lobe filled
planet. Also, the code ignores thermal properties of the solid core of a giant
planet, leaving a question as to what would happen to the core under the intense
tidal heating. In a hot Jupiter-star system, the mass loss may occur via the
Lagrangian 2 point as well \citep{Gu03}, which is not considered in the present
work. Besides, the mass loss from a planet in an eccentric orbit may be
intermittent and nonconservative, while the mass loss rate in our calculation is
simply estimated from the L1 overflow at $r=a$. To use a more appropriate
expression for the Roche radius and to consider the more realistic mass loss in
asynchronous eccentric planetary systems, a more careful treatment
\citep[e.g.][]{Sepinsky} is desired to refine the results. In light of these
limitations and possible uncertainties, the simulations covering a wide range of
the parameter space along with more realistic modelling on individual issues will
be explored in a future work.



\acknowledgments

We are grateful to G. I. Ogilvie, F. Pont, and R. E. Taam for useful
discussions. This work has been supported by the NSC grant in Taiwan through NSC
98-2112-M-001-011-MY2.

\appendix
\section{Appendix}
The Roche radius $R_l$ is defined as the distance between L1 and the planet's
center of mass. The conventional expression of $R_l$ is derived from the circular
restricted three-body problem \citep[e.g.][]{Murray1999}. \citet{Gu03} assumed that
the Roche lobe overflow occurs around perihelion and made an approximation that
$R_l=( M_p/3M_* )^{1/3} a(1-e)$. Here we investigate the validity of the
approximation by employing the expression for the L1 point in the elliptical
restricted three-body problem \citep{Todoran93}. The exact values of the Roche
radius in units of $a$ are calculated using the equations in \citet{Todoran93} and
are shown in Table \ref{tbl-1} for various mass ratios and eccentricities relevant
to our problem. The table shows that the Roche radius at perihelion is larger than
that at aphelion in some of the cases. It is because the planet orbits faster
(slower) than the ``fictitious" circular motion performed at perihelion (aphelion).
As a result, the centrifugal force due to the orbital motion is larger (smaller) at
perihelion (aphelion) than the centrifugal force of the fictitious circular motion
at the point, therefore enlarging (reducing) the Roche radius relative to the
``Roche radius" associated with the circular motion.

Furthermore, the above analysis is based on the assumption that the spin of the
planet is instantaneously synchronized with the orbit at any moment \citep{Pratt}.
While it is true for a synchronized circular orbit, in an eccentric orbit the
planet's spin is normally ``synchronized" with its orbital motion close to
perihelion where the tidal forcing is expected to be near its maximum
\citep{Hut81,IP07,LL08}. Following the same line of the above argument, a
``synchronized"  planet spins faster than its orbit at aphelion, implying that the
Roche radius is smaller than the value associated with the circular motion at the
point.

It should be noted that equation (\ref{eq:M_dot}) for the mass loss rate is also derived
for a circular orbit. The equation should be modified as the L1 point changes its
location along an eccentric orbit. Besides, when the planet's photosphere is close
to its time-varying Roche lobe in an eccentric orbit, the outer part of the planet
should expand or contract with the Roche lobe on the local dynamical timescale,
which introduces additional complexity to estimate the mass loss rate.

It is for the above complex reasons that we simply use the expression
$R_l=(M_p/3M_*)^{1/3}a$ to specify the Roche radius of a planet in an eccentric
orbit in the present work.

\begin{table}
\begin{center}
\caption{Roche radius in units of $a$ calculated for different mass ratios and
eccentricities \label{tbl-1}} \

\begin{tabular}{crrrrrrrrrrr}
\tableline\tableline $M_p/M_\ast$ & $e$ & Roche radius at perihelion $(a)$ &
Roche radius at aphelion $(a)$  \\
\tableline
$2\times10^{-3}$ &0.1 &0.0851 &0.0842\\
$2\times10^{-3}$ &0.3 &0.0831 &0.0833\\
$1\times10^{-4}$ &0.1 &0.0411 &0.0265\\
$1\times10^{-4}$ &0.3 &0.061 &0.0221\\
\tableline\tableline
\end{tabular}
\end{center}
\end{table}




\clearpage



\clearpage

\clearpage

\clearpage


\begin{figure}
$$
\scalebox{0.5}{\includegraphics{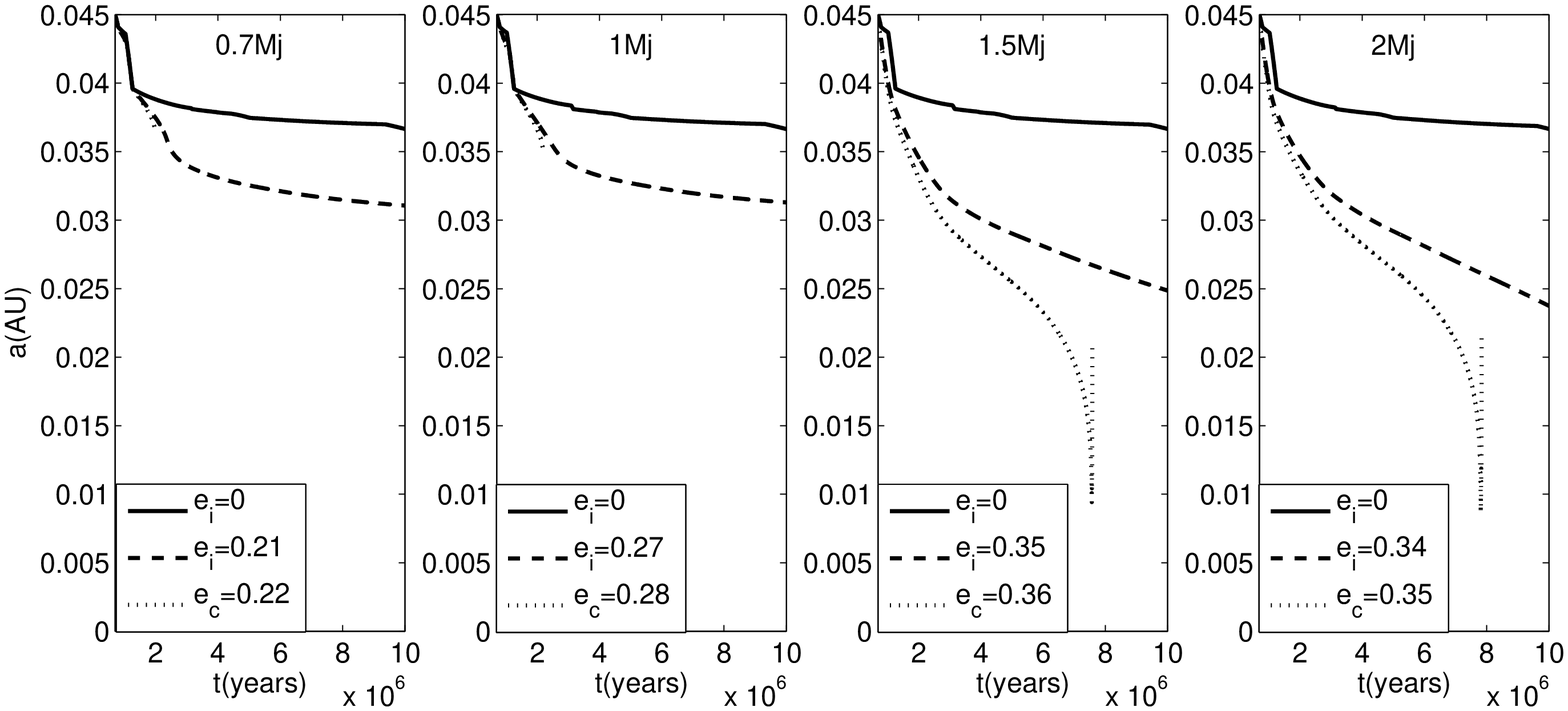}}
$$
\caption{Orbital evolutions of the planets in the large cavities. By modulating the
initial eccentricity, the orbital evolution will change. Each panel displays the
results for a different initial $M_p$. The solid line demonstrates the orbital
evolution of the planet with $e_i=0$; the dashed line illustrates the case in
which the planet of a given initial mass can migrate the longest distance; the dotted
line shows that the planet with $e_i=e_c$ destructs at the breaking point.}
\label{bigcavity}
\end{figure}

\clearpage



\begin{figure}
$$
\scalebox{0.5}{\includegraphics{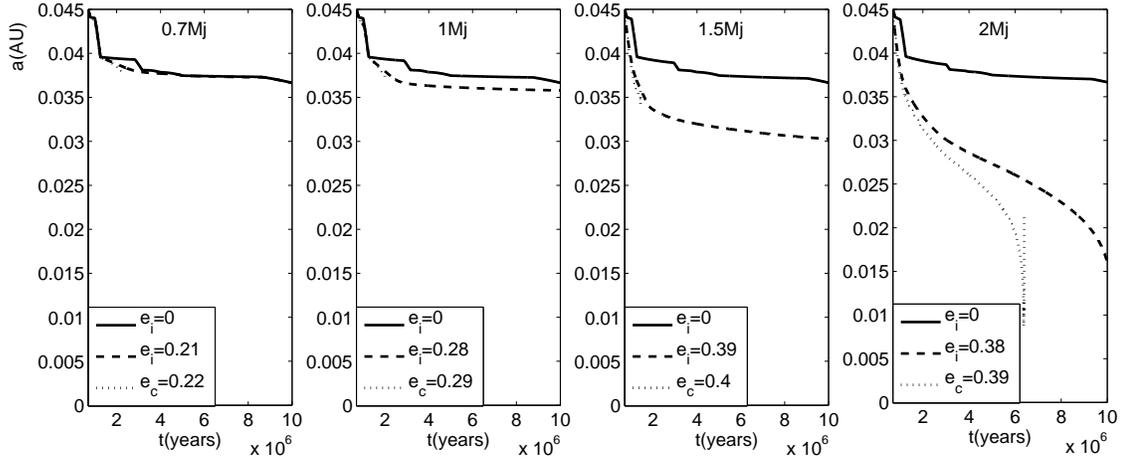}}
$$
\caption{The same as Figure \ref{bigcavity} except that the coefficient $\epsilon$ is
reduced to 0, meaning that the magnetic fields do not exert any torque on the
planet. The critical eccentricities $e_c$ are relatively high compared to the
previous cases for $\epsilon=1$.} \label{eps0}
\end{figure}

\clearpage


\clearpage





\end{document}